\documentclass[pdflatex,sn-basic,iicol]{sn-jnl}

\usepackage{graphicx}%
\usepackage{multirow}%
\usepackage{amsmath,amssymb,amsfonts}%
\usepackage{amsthm}%
\usepackage{mathrsfs}%
\usepackage[title]{appendix}%
\usepackage{xcolor}%
\usepackage{textcomp}%
\usepackage{manyfoot}%
\usepackage{booktabs}%
\usepackage{algorithm}%
\usepackage{algorithmicx}%
\usepackage{algpseudocode}%
\usepackage{listings}%
\usepackage{multirow}
\usepackage{boxedminipage}

\theoremstyle{thmstyleone}%
\theoremstyle{thmstyletwo}%

\theoremstyle{thmstylethree}%

\begin{document}

\title[Search Timelines]{Search Timelines: Visualizing Search History to Enable Cross-Session Exploratory Search}


\author*[1]{\fnm{Orland} \sur{Hoeber}}\email{orland.hoeber@uregina.ca}

\author[1]{\fnm{Md Nazmul} \sur{Islam}}\email{mdnazmul@uregina.ca}

\author[1]{\fnm{Miriam} \sur{Boon}}\email{miriam.boon@uregina.ca}

\author[2]{\fnm{Dale} \sur{Storie}}\email{dale.storie@uregina.ca}

\author[2]{\fnm{Veronica} \sur{Ramshaw}}\email{veronica.ramshaw@uregina.ca}

\affil*[1]{\orgdiv{Department of Computer Science}, \orgname{University of Regina}, \orgaddress{\street{3737 Wascana Parkway}, \city{Regina}, \postcode{S4S 0A2}, \state{SK}, \country{Canada}}}

\affil[2]{\orgdiv{Dr. John Archer Library}, \orgname{University of Regina}, \orgaddress{\street{3737 Wascana Parkway}, \city{Regina}, \postcode{S4S 0A2}, \state{SK}, \country{Canada}}}

\abstract{\textbf{Purpose:} The timespan over which exploratory searching can occur, as well as the scope and volume of the search activities undertaken, can make it difficult for searchers to remember key details about their search activities. These difficulties are present both in the midst of searching as well as when resuming a search that spans multiple sessions. In this paper, we present a search interface design and prototype implementation to support cross-session exploratory search in a public digital library context.
\textbf{Methods:} Search Timelines provides a visualization of current and past search activities via a dynamic timeline of the search activity (queries and saved resources). This timeline is presented at two levels of detail. An Overview Timeline is provided alongside the search results in a typical search engine results page design. A Detailed Timeline is provided in the workspace, where searchers can review the history of their search activities and their saved resources. A controlled laboratory study (n=32) was conducted to compare this approach to a baseline interface modelled after a typical public digital library search/workspace interface.
\textbf{Results:} Participants who used Search Timelines reported higher levels of user engagement, usability, and perceived knowledge gain, during an initial search session and when resuming the search after a 7-8 day interval. This came at the expense of the searchers taking more time to complete the search task, which we view as positive evidence of engagement in cross-session exploratory search processes. 
\textbf{Conclusion:} Search Timelines serves as an example of how lightweight visualization approaches can be used to enhance typical search interface designs to support exploratory search. The results highlight the value of providing persistent representations of past search activities within the search interface.}




\maketitle

\bmhead{Statements and Declarations}
Competing Interest: This work was supported by the Natural Sciences and Engineering Research Council of Canada (NSERC), through the Alliance Grant (ALLRP 558319-20) held by the first author, in collaboration with the fourth and fifth authors, the Regina Public Library, and the Saskatchewan Information and Library Services Consortium.

\section{Introduction}\label{sec1}

A public digital library is an online platform that provides access to a wide range of library resources (including books, multimedia materials, and other educational content) to the general public~\citep{leckie_understanding_2005, ma2006digital}. They allow library patrons to search for and access library holdings without the constraints of library hours or physical locations. The physical libraries often have multiple branches within a municipality, and may be part of a regional consortium to provide economies of scale to their digital library services. 

The search interfaces provided by the public digital library follow common design patterns that are prevalent across many digital library contexts (e.g., a query box and a list of search results). When searching for known or easily described items, the simple search interfaces provided by public digital libraries can help library patrons find the resources they are seeking. However, such simple interfaces provide limited support to searchers when their tasks become complex~\citep{hoeber-2025}.

The information needs of searchers in public digital libraries may be resolved through search behaviour that ranges from simple to complex~\citep{belkin1995cases}. Searchers engaging in simple lookup searches have a clear understanding of their information needs and can generate well-defined queries to retrieve relevant documents~\citep{Marchionini2006}. However, library searchers may also need to undertake complex search tasks that are ill-defined, open-ended, or multifaceted, and cannot be resolved by finding a single best document~\citep{White2009}. 

While undertaking complex search tasks, simple lookup approaches are not sufficient~\citep{belkin1995cases, belkin2004evaluating}; instead, searchers may engage in a process of exploratory search~\citep{White2009}. Exploratory search consists of two high-level types of searching: exploratory browsing and focused searching. Searchers start with an exploratory browsing style of search, and as their uncertainty in the task is reduced, they transition to focused searching. Throughout both stages of exploratory search, searchers issue many queries and save many resources, yet typical public digital library search interfaces provide few services to help searchers keep track of what they are doing beyond a simple workspace for saving search results.


To resolve complex search tasks, searchers may undertake searches that span multiple sessions over an extended period of time~\citep{Agichtein12, Capra20}. That is, they may start a search session and perform some initial (exploratory browsing) searching to familiarize themselves with what is available. As they discover new information, they may wish to pause their searching to give themselves an opportunity to think about what was found. Or they may have time constraints that keep them from finishing the task.

To resume and reacquaint themselves with the previous search activities, searchers need explicit support from the search interface~\citep{Morris08, Capra20, Capra21}. Unfortunately, while performing cross-session searching, searchers are not well supported by the majority of public digital library search interfaces, as little information is available to help searchers reacquaint themselves with their past search activities beyond a simple list of saved resources in a workspace. Because of this, searchers are forced to reacquaint themselves and resume the past search tasks on their own, and must come up with individual strategies for carrying out cross-session searches~\citep{Capra20, Capra21}.

Keeping track of past search activity is a fundamental challenge in exploratory search due to the active nature of such searching and the need to issue multiple queries in the exploratory search process. Re-finding previously viewed documents or reissuing previously searched queries is common while performing complex search tasks~\citep{Marchionini2006, White2009}. However, this may not be easy when the search tasks extend over a prolonged period of time~\citep{Agichtein12, Capra20, Morris08, Capra21}. As a result, in this research we focus on the problem space that is at the intersection of exploratory search and cross-session search: cross-session exploratory search.


A key issue to be addressed in cross-session exploratory search is the lack of search interface support for helping searchers to keep track of their search activities. While many digital library search interfaces provide support for saving resources within workspaces, much of the context for why a particular resource was saved is lost. Further, if a searcher wants to review what they have found thus far, they must leave the search interface and view the workspace, which may make it difficult to get back to where they were in their search process. 

With these problems in mind, we have designed and implemented a public digital library search interface we call \textit{Search Timelines}. The overall approach is to provide visual timelines of the search activities at two levels of detail. An Overview Timeline is provided as a lightweight addition to the typical search engine results page, which allows searchers to keep track of their queries and saved resources without having to leave the search interface to view what they have done thus far. A Detailed Timeline is provided in the workspace, where all saved resources are presented in the same format as was used in the search results list, and are organized in the context of when they were saved and as a result of which query. This organizational structure is designed to allow searchers to readily reacquaint themselves with the progress they made in past search sessions as they resume their search.

The remainder of this paper is organized as follows. A literature review is provided in Section 2, establishing the foundations for this research. A detailed description of the design of Search Timelines is provided in Section 3. The methods for evaluating this approach in comparison to a typical baseline system are provided in Section 4. The results of the evaluation are provided in Section 5, followed by a discussion on how these findings relate to previous work and what new knowledge has been discovered in Section 6. The paper concludes with a summary of the primary contributions and an overview of future work in Section~7.

\section{Literature Review}\label{sec2}

While searching within public digital libraries is often focused on known item lookup  searches that can be completed in a single search session~\citep{belkin1995cases}, sometimes search tasks are so complex that the search must extend into multiple search sessions~\citep{Capra21}. Such complex multi-session searches occur because there is a high level of cognitive processing that a searcher must perform as they seek to resolve their information needs~\citep{Capra21}, making it difficult to complete the task in a single session.

When undertaking a complex search task that requires higher-level cognitive processing, simple models of search behaviour are not sufficient. Instead, many consider White \& Roth's exploratory search model~\citep{White2009} to be a useful framework to describe how to undertake such searches~\citep{soufan-2022,Gomes-Hoeber-22,hoeber22}. This model of exploratory search describes two different types of searching that one might undertake when faced with a complex search task: exploratory browsing and focused searching. When performing exploratory browsing style of searching, the focus is on discovering, learning, and investigating behaviours. When sufficient information has been found to reduce uncertainty in the complex search task, a transition is made to focused searching. Here, the focus is on reformulating queries to isolate specific information, examining the search results in detail, and extracting information.

Given that exploratory search is a model that is well-suited to the pursuit of complex search tasks, and such complex search tasks may extend across multiple search sessions~\citep{Capra20, Morris08, Han15}, there is a synergy between exploratory search and cross-session search~\citep{Gomes-Hoeber-22}. This assertion is consistent with the work on ``transmuting successive searches''~\citep{Lin13}, where searchers learn about and refine their information needs over multi-session searches. 

A recent study explored the reasons searchers stop a search and found that nearly half were related to having to synthesize the information found~\citep{Capra20}, with implications for continuing the search later. This same study also found that there are many external factors at play that can also cause one to stop their search activity, including time limitations, fatigue, and distractions~\citep{Capra20}. 

Returning to a search task is cognitively taxing, as searchers often need to rely on their memory to resume and pick up the search where they left off~\citep{Capra20, Capra21, Morris08, Han15}. A critical first step is to reacquaint ones self with the prior work, which serves as a bridge between what has been done already and what will be done now~\citep{Gomes-Hoeber-22}. A typical starting point for continuing a previous search is to undertake a re-finding style of search~\citep{Han17}. Common techniques include reviewing saved information and notes taken about these, as well as re-issuing previous queries~\citep{Capra20, MacKay08, Morris08}. While some searchers have developed clear strategies for search task resumption, many have difficulty when asked to articulate what they do to reacquainting themselves with their prior search~\citep{Capra20}. 

A typical approach to supporting cross-session search within an interface is to include a workspace where resources can be saved. Although many digital library search interfaces include rudimentary workspaces, there are also recent studies on how to enhance workspaces to make them more useful for exploratory search~\citep{orgbox21, di_sciascio18, hoeber22}. Such approaches focus on helping a searcher to make sense of \textit{what} search results were saved, but not necessarily \textit{how} they were found.

Grouping saved information by tasks or sessions can make search resumption easier; providing visual representations of such information can help as well~\citep{Gomes-Hoeber-22}. Whether such groups should be created automatically~\citep{Abela16, Xu18, Morris08} or explicitly~\citep{Buivys16, MacKay09, Bento2018, Gomes-Hoeber-22} remains an open question. While automatic methods allow the searcher to avoid extra barriers to starting a search, explicit methods put them in a position of control of their search activity.

We can also consider workspaces from the perspective of how they support searchers in contextualizing the found information. Such contextualization can be divided into two categories: metadata about the search results themselves, and metadata about the search process~\citep{golovchinsky_future_2012}. Metadata about search results typically consist of details such as keywords, topics, facets, and entities extracted from the documents~\citep{Gomes-Hoeber-22, hoeber22, chang_searchlens_2019}, and may also include snippets~\citep{Xu18, Qvarfordt14}, or aspects selected or created by the user such as notes or annotations~\citep{orgbox21}. Metadata about the search process organizes the information based on details such as which search results the user has viewed, how many times a particular search result has been returned by the user's searches~\citep{Qvarfordt14}, and temporal details of the search activities \citep{Gomes-Hoeber-22, Xu18, Hoeber09}. 

A variety of mechanisms have been developed for presenting search topic/session information to searchers. These include graph-based visualizations~\citep{Abela16, Xu18}, colour-encoding of saved resources~\citep{Qvarfordt14},  timelines~\citep{Hoeber09, Xu18, Gomes-Hoeber-22}, and hierarchical representations~\citep{Morris08}. 

Of particular relevance to this research are the timeline used in LogCanvas~\citep{Xu18} and the hierarchical representation used in SearchBar~\citep{Morris08}. Both approaches provide mechanisms for evaluating past search activities, but go about it in different ways. LogCanvas uses a vertical timeline to represent search sessions. For each search session, the start and end time are specified, along with a list of all queries that were issued during the session. This approach allows a searcher to see when queries were previously issued, but without an ability to assess their success in surfacing relevant search results. SearchBar uses a collapsible tree structure representation of search topics, queries issued, and search results viewed. This approach allows a searcher to see which queries lead to relevant search results (those the searcher chose to view or mark as useful), but without an ability to see when the queries were issued or when individual search results were viewed.

\section{Design of Search Timelines}\label{sec3}

The design of Search Timelines was inspired by prior work on presenting search topics and sessions to searchers. Drawing from LogCanvas~\citep{Xu18}, it provides a vertical timeline of the search sessions. Drawing from SearchBar~\citep{Morris08}, it provides both queries and saved search results in the timeline. By combining features from these two approaches, we address their respective limitations noted in the previous section. While both of these prior approaches were developed to evaluate past search activities, a novel aspect of Search Timelines is its ability to document current search activity through integration within the search engine results page (SERP). Considering the Design Principles for Exploratory Search Interfaces~\citep{hoeber-2025}, Search Timelines fulfills all five principles: lightweight additions, visual, scrutable, interactive, and persistent. In the remainder of this section, the key features of Search Timelines are described in detail.

\subsection{Search Topic Management}


Search topics represent the different search interests a user wishes to keep separate from one another. A fundamental principle in search topic management is that each search topic must preserve a separate history and list of saved search results. To support this, we have implemented a simple workspace approach that follows a pattern that has been shown to be effective in other approaches~\citep{Gomes-Hoeber-22}. When beginning a new session, the searcher’s first query becomes the topic title for the workspace. If the searcher finds at some point that this is no longer an accurate description of the topic, it can be edited from within the workspace.  

Using the concept of an ongoing (current) search topic, all user actions including issuing new queries, saving search results, and removing saved search results, occur within the context of the ongoing search topic's workspace. This ongoing topic also determines what information to display in the timeline visualizations. A list of all available workspaces is also provided, enabling the searcher to easily switch between search topics, as well as create new ones

\subsection{SERP and Overview Timeline}

The SERP follows a typical digital library search interface design pattern. Each standard search result card includes typical resource information (book cover, title, author, year of publication) along with a button that allows the searcher to save that search result to the ongoing topic's workspace, or remove it from the workspace if it has already been saved.

This interface is augmented with an Overview Timeline provided to the right of the search results list. This timeline is automatically generated and updated in real-time as users engage in search activities, reflecting the interconnectedness of these actions within the search context. 

\begin{figure*}[ht]
  \begin{minipage}[t]{0.95\linewidth}
    \centering
     \frame{\includegraphics[width=\linewidth]{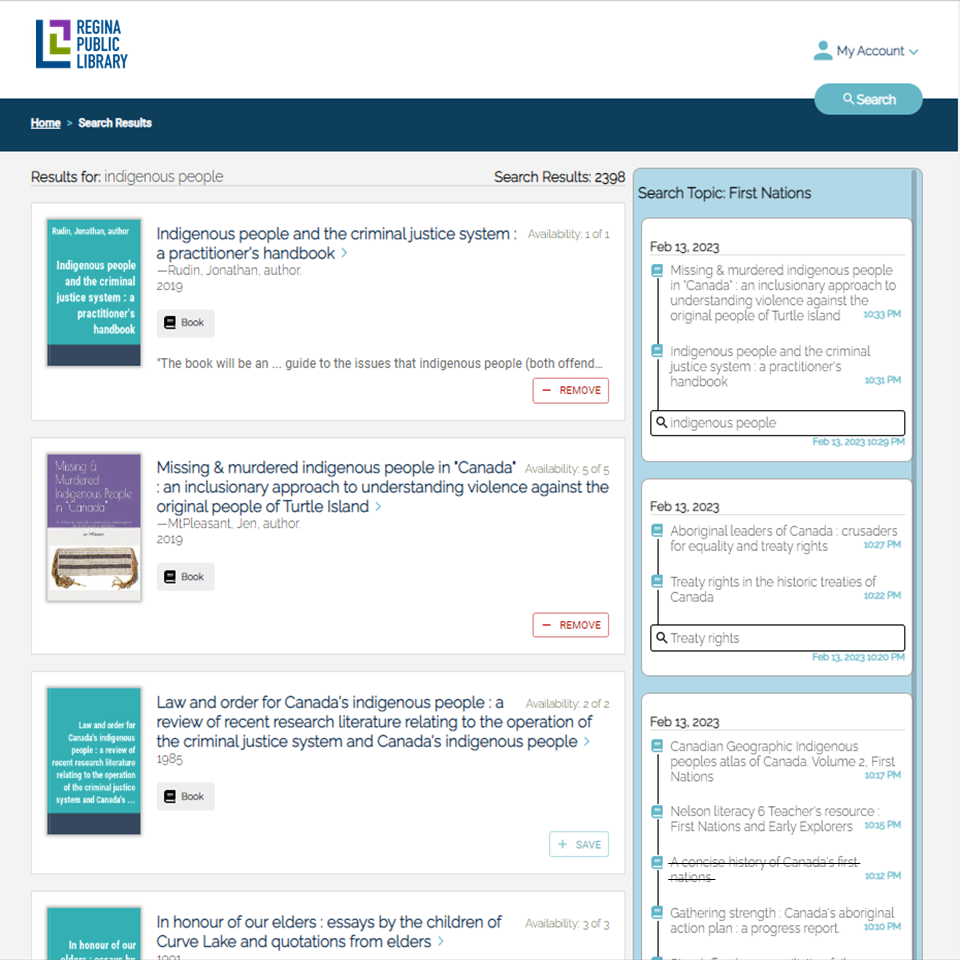}}
    \caption{The SERP view of Search Timelines, displaying results for the query, ``indigenous people''. Two documents from this query have already been added to the ongoing search topic, ``First Nations'', and can be seen in the Overview Timeline to the right. The Overview Timeline also shows information from two earlier search sessions.}
    \label{fig:SERP}
  \end{minipage}\hfill
\end{figure*}

As a query is issued, it is added to the timeline with a timestamp specifying the date and time the query was issued. Search results that are saved to the workspace from this query's SERP are added to the timeline using an icon that represents the resource type (e.g., book, DVD, etc.), the title of the search result, and the time it was saved (see Figure \ref{fig:SERP}). Bounding boxes encapsulate a query and the search results saved from that query, with explicit lines connecting the icons of the saved search results back to the query. These approaches reinforce the relationships between saved search results and queries, following the Gestalt principles of enclosure and connectedness \citep{koffka-1935, gestalt}. Structurally, the timeline adopts an upside-down tree-like configuration, where the most recent activities are displayed at the top while evidence of older search activity descend downwards.

The benefit of providing the Overview Timeline within the SERP is that it enables a searcher to monitor how their search activity is progressing without having to leave the primary search interface. A quick visual scan of the timeline can reveal how many queries have been issued recently, which have been particularly valuable in finding relevant search results, and what types of resources have been saved. Doing so can provide confidence in progress being made in resolving the information need, which contributes to a reduction in the uncertainty associated with undertaking complex search tasks. Further, reviewing the titles of the saved search results can reveal textual patterns that contribute to learning about the topic and the development of new queries on different facets of the search task. 

One potential limitation of the Overview Timeline is that it could grow very large for search tasks that extend over long periods of time and include many queries and saved resources. Although the solutions to this problem are beyond the scope of this research, it could be addressed by limiting the number of sessions shown or providing an interactive collapsing and expanding of the sessions with the older sessions being collapsed by default.


The value of this Overview Timeline is particularly evident in the prolonged nature of exploratory searches. Searchers often engage in iterative processes, navigating back and forth between past queries and saved documents~\citep{White2009, Morris08, Teevan10, Dan2018:2}. The timeline serves as a visual aid, enabling searchers to see earlier search actions that have been undertaken during the exploratory search process, rather than having to remember such information. In particular, it can enable searchers to easily see what facets of their complex search tasks have been investigated thus far, enabling the searcher to plan their next course of action while undertaking an exploratory search process.



\begin{figure*}[ht]
  \begin{minipage}[t]{0.95\linewidth}
    \centering
     \frame{\includegraphics[width=\linewidth]{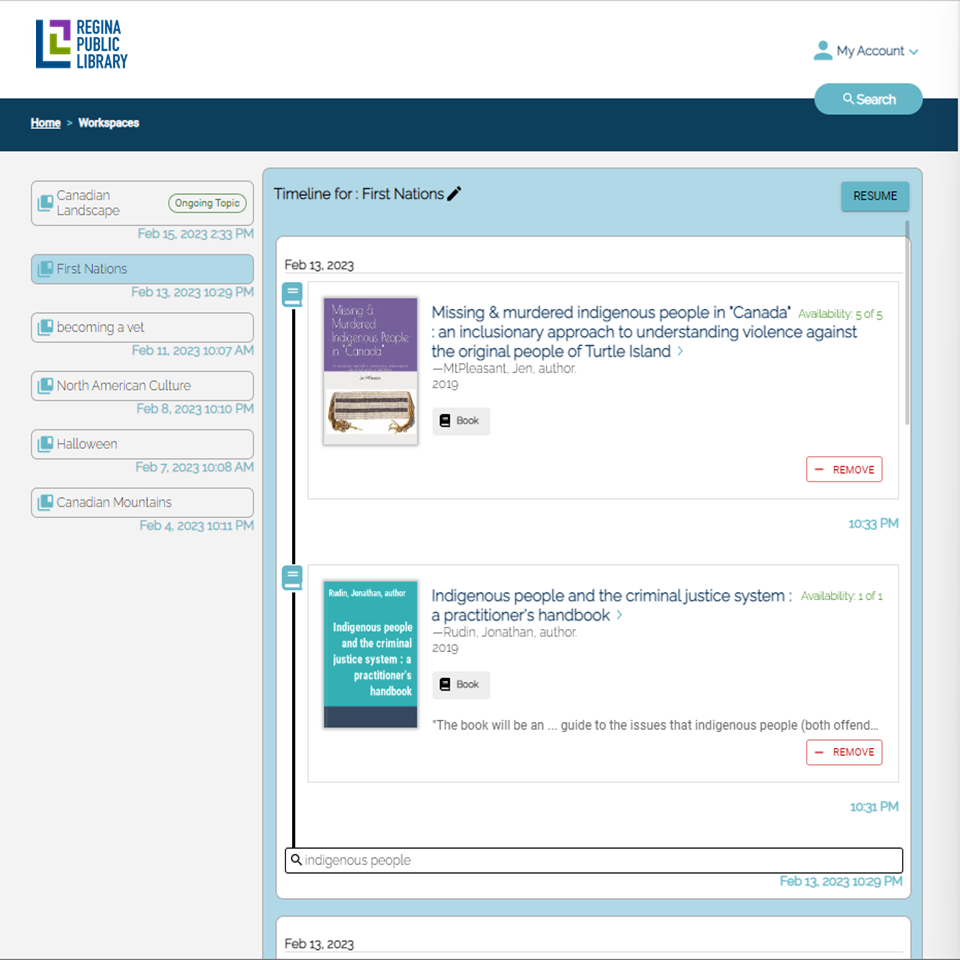}}
    \caption{The workspace view of Search Timelines, with ``First Nations'' selected as the displayed topic along with the Detailed Timeline of the search activities for this topic. The most recent search session is shown, including the query and saved resources.}
    \label{fig:Workspace}
  \end{minipage}
\end{figure*}

\subsection{Workspace and Detailed Timeline}

Information saved to the workspace is organized as a Detailed Timeline, which merges the design of the overview timeline with the design of the search results cards from the SERP (see Figure \ref{fig:Workspace}). The mechanism for linking the icons of the saved search results back to their source queries is the same between both timelines, although here the icons are larger and the lines connecting them to the queries are thicker and darker. Rather than only listing the title of the saved search result (as is done in the Overview Timeline), the complete search result card is used.

The workspace can be used in the context of an ongoing search topic, providing searchers with the ability to assess the details of the saved search results. This an important feature in the context of an exploratory search process since the searchers' understanding of their topics will change as they discover new information and learn about the topic~\citep{White2009}. Any previously saved search results that are determined to no longer be relevant can be removed from the workspace by clicking on the ``remove'' button. In order to preserve the timeline of the search activity, this does not completely delete the item, but instead uses a strike-through on the title and reduces the saturation of the font and icons to minimize the visual presence of these search results. This occurs for both the Detailed Timeline and the Overview Timeline.

A key benefit of the Detailed Timeline represented in the workspace is that it can help searchers to reacquaint themselves with past search activities when returning to a search topic that was not completed in a single session. Such support is critical in enabling cross-session searching~\citep{Gomes-Hoeber-22, Capra20, Capra21}. When returning to their chosen topic, searchers can quickly assess the search results that were saved and the queries that were used to get to these search results. They can be examined in detail to remind the searchers why they were saved in the first place. The searchers can see what was considered and then rejected. They can choose to re-issue any previous query simply by clicking on it, or may issue a new query in the context of the search topic by clicking on the search button at the top of the page (which expands to reveal a query panel). In either case, the resulting SERP loads the Overview Timeline, enabling the searchers to see their resumed search activity in the context of what they had done previously.

When navigating to the workspace, the default is to load the workspace for the ongoing topic (represented by an ``ongoing topic'' badge alongside the topic name). A list of the previous topics is provided, ordered by the date and time of the most recent activity on the topic. Switching between topics is done by simply clicking on the topic title; doing so updates the Detailed Timeline and uses colour encoding to show the correspondence between the selected topic and the timeline (both have a blue background colour). A resume button is provided to allow the searcher to make the selected topic the current (ongoing) topic. 

\begin{figure*}[ht]
  \begin{minipage}[t]{0.95\linewidth}
    \centering
     \frame{\includegraphics[width=\linewidth]{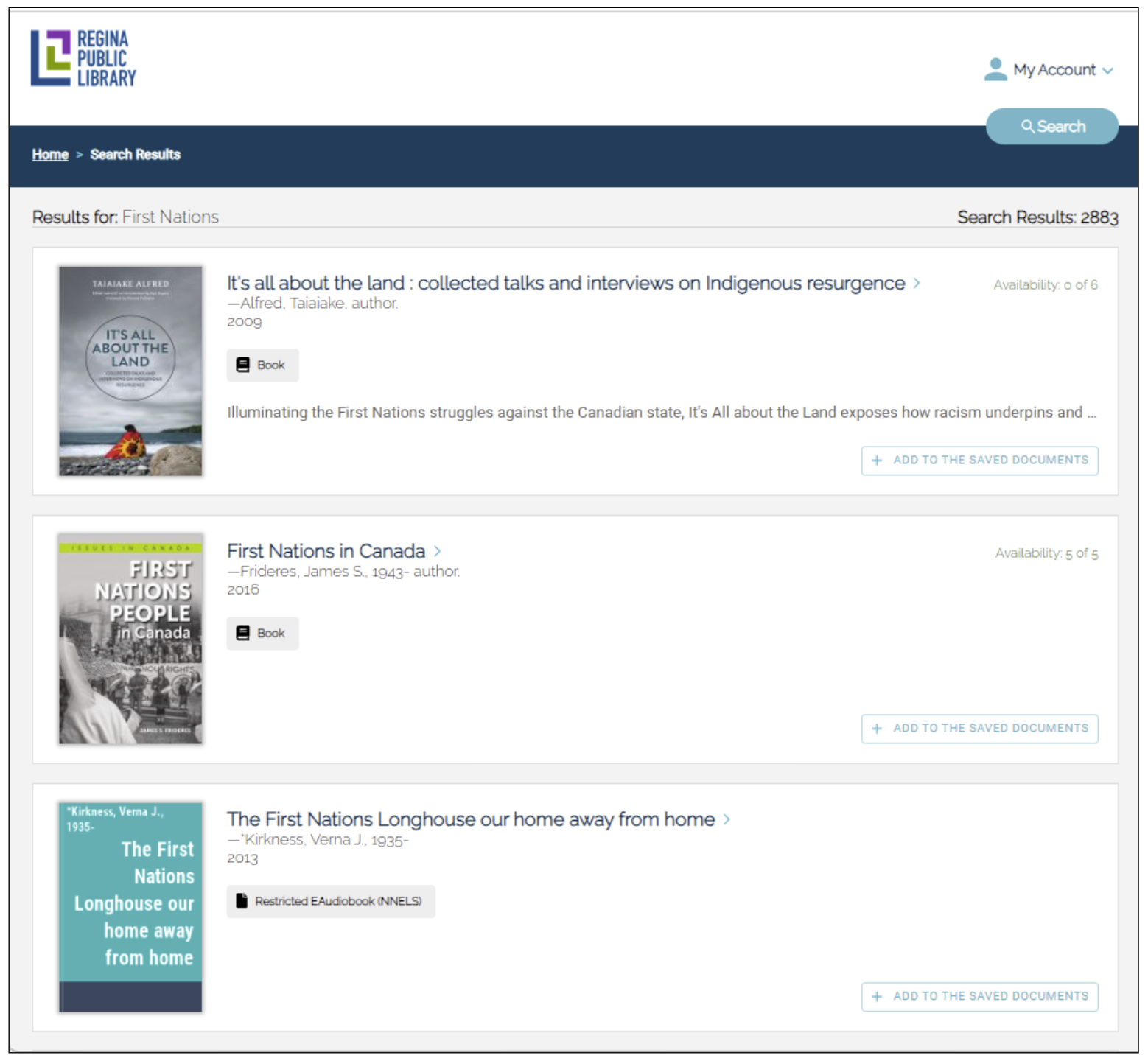}}
    \caption{The Baseline interface uses a common digital library search design pattern, displaying results for the query ``First Nations''. Resources can be saved to the workspace with the ``add to saved documents button''. The workspace uses a similar visual representation for the saved resources. }
    \label{fig:Baseline}
  \end{minipage}
\end{figure*}

\subsection{Prototype Implementation}
The Search Timelines prototype was implemented using the MERN stack (MongoDB, Express, React, Node), and the MaterialUI library. Search queries are submitted to and processed by a RESTful API managed by a regional public library consortium, returning a ranked list of search results in JSON format. Each user's workspaces, queries, and saved resources are stored in a centralized database. 

\section{Evaluation}\label{sec4}

In order to evaluate whether Search Timelines is effective at enabling cross-session exploratory search within a public digital library context, we conducted a user study comparing Search Timelines to a baseline search interface which is modelled after a typical library search interface (see Figure \ref{fig:Baseline}). The design of the study was motivated by the following eight research questions, which are provided in pairs (initial search and resumed session):

\textbf{RQ1a:} What is the difference in user engagement when using Search Timelines versus the Baseline for conducting digital library searches?
        
\textbf{RQ1b:} What is the difference in user engagement when using Search Timelines versus the Baseline for resuming a search task?
        
\textbf{RQ2a:} What is the difference in the perception of usability (ease-of-use, usefulness, and satisfaction) when using Search Timelines compared to the Baseline for conducting digital library searches?

\textbf{RQ2b:}  What is the difference in the perception of usability (ease-of-use, usefulness, and satisfaction) when using Search Timelines compared to the Baseline for resuming a search task?

\textbf{RQ3a:} How is perceived knowledge gain affected when using Search Timelines compared to the Baseline?
        
\textbf{RQ3b:} How is perceived knowledge gain affected when using Search Timelines compared to the Baseline when resuming the task?

\textbf{RQ4a:} What is the difference in the time to task completion when using Search Timelines versus the Baseline for conducting digital library searches?
        
\textbf{RQ4b:} What is the difference in the time to task completion when using Search Timelines versus the Baseline for resuming the search task?
        
Because both interfaces used the same underlying search results provided by our local public library, and presented the search results in similar styles on the SERP, we did not expect any difference in the ability for searchers to find and save relevant search results. As such, we did not pose a research question related to the precision of the saved search results.

\subsection{Methodology}
A between-subjects controlled laboratory study design~\citep{Kelly2009} was used to evaluate the search interface, with a longitudinal component spanning two sessions. The inclusion of this longitudinal aspect was essential to capture the dynamics of cross-session searching. To avoid introducing a confounding factor related to topic/workspace switching, a between-subjects design allowed the use of a single search topic. Moreover, this study allowed for an evaluation of how search behaviour might be influenced by extended gaps between sessions and how searchers approached the resumption of a previously unfinished task.

A single independent variable was used in this study: the search interface assigned to the participants (Search Timelines or Baseline). All other aspects of the study were kept constant to allow for potential causal relationships to be inferred.

For addressing research questions RQ1a and RQ1b, we measured user engagement, which is a pivotal aspect of the overall user experience~\citep{oBrian2016}. To do so, we employed the User Engagement Scale (Short Form)~\citep{OBrien18}, which is composed of four dimensions: focused attention (FA), perceived usability (PU), aesthetic appeal (AE), and reward factor (RW).
           
In order to address research questions RQ2a and RQ2b, three key measures of usability were selected: perceived ease-of-use, perceived usefulness, and perceived satisfaction. These collectively offer a comprehensive view of perceived usability~\citep{Kelly2009, venkatesh2000theoretical, pearson1980measurement}. For perceived ease-of-use and perceived usefulness, the questions were structured based on the Technology Acceptance Model 2 (TAM2)~\citep{venkatesh2000theoretical}. Perceived satisfaction was measured using a set of questions that were modeled after the TAM2 instrument, following an approach that has been used in other studies~\citep{hoeber22}.
    
Research questions RQ3a and RQ3b deal with the concept of perceived knowledge gain. This variable is concerned with gauging the extent to which searchers believe their comprehension and awareness of the task has either expanded or diminished throughout their interaction with the system~\citep{White2009}. To measure this concept, we used three questions focused on perceived knowledge of the task (level of current knowledge, ease of explaining the task to others, and confidence in searching for information on the topic), administered both before the search task was started and after it was completed for the given session. Each of these was measured using five-point Likert scale, with the difference representing the change in perceived knowledge gain. Perceived knowledge gain reflects the cognitive shift experienced by users during their interaction with the system, contributing to a comprehensive assessment of the overall impact on users' information-seeking capabilities and task comprehension~\citep{soufan-2022}.
 
To address research questions RQ4a and RQ4b, measurements were made for the total duration users spent utilizing the interface for their search activities~\citep{Kelly2009}. Such measurements were made independently for the initial and resumed search sessions. The data for this was extracted from the search activity logs.

\subsection{Study Procedures}
\begin{figure}[h]
  \centering
  \frame{\includegraphics[width=1\columnwidth, scale=2]{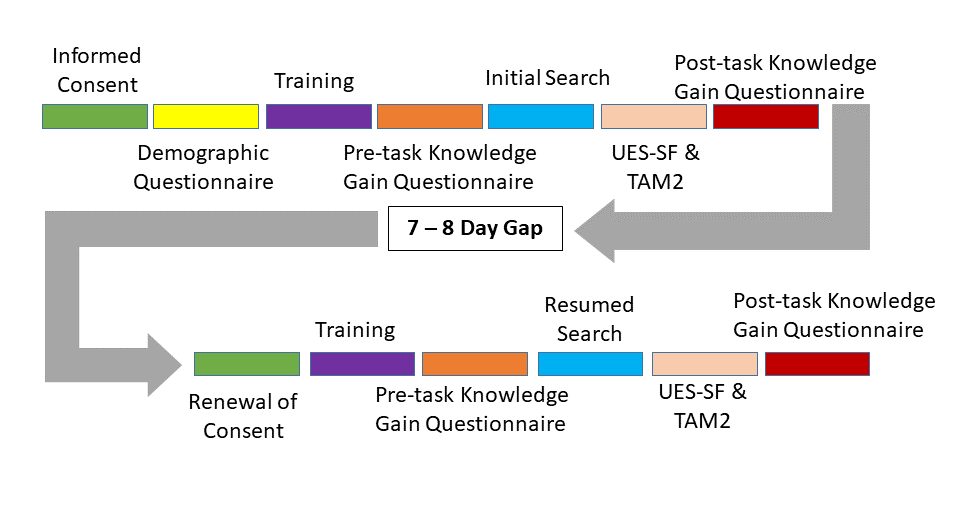}}
    \caption{Structure of the user study that shows each step for both initial and resumed session between 7-8 day gap.}
    \label{fig:study}
\end{figure}

Each participant followed the same study procedures, with the only difference being that half the participants were given Search Timelines to use while conducting the search task, while the other half were given the Baseline. In order to help participants familiarize themselves with the assigned search interface, a pre-recorded training video was shown to the participants and they were then given an opportunity to use the interface to recreate the search scenario shown in the training video. A similar training session (video and experiential training) was provided for the second phase of the study, focusing on resuming the previous training scenario task. Other details of the study procedures are illustrated in Figure~\ref{fig:study}. The study was conducted online using Zoom, with screen sharing activated so that participant progress through the steps of the study could be monitored. A Qualtrics survey was used to provide links to the assigned interface, the training material, and to collect the survey data.

Of particular importance in this study design is the creation of an environment where the task cannot be completed in a single search session. To do so, we limited the initial search to a maximum of 15 minutes and scheduled the second search session to occur 7-8 days later. This time gap between sessions is consistent with other studies on cross-session search~\citep{Morris08, Capra10, Gomes-Hoeber-22}. The second search session had no time limit. The rest of the study followed a typical interactive information retrieval user study pattern. The entire process was reviewed and approved by the Research Ethics Board at our university.



\subsection{Simulated Search Task}

The concept of `simulated work tasks' includes the source of the information need, the situational context, and the specific problem to be solved~\citep{Borlund97}. In this research, the simulated work task revolves around the process of gathering documents from a digital public library for the purpose of composing a high school class assignment on a history topic (see Figure~\ref{fig:tasks}). This task was designed thoughtfully in consultation with librarians to ensure sufficient coverage of the topic in the library holdings.

\begin{figure}[!t]
\small
\begin{boxedminipage}{\columnwidth}
    
\textbf{Search Scenario:} Suppose you are a high school student and taking a class on History. You have been given a class assignment to write a brief report on the French Revolution. The French Revolution was a period of political and social upheaval in France from 1789 to 1799. The Revolution was triggered by a combination of factors, including a financial crisis, food shortages, and widespread dissatisfaction with the absolute monarchy and the privileges of the aristocracy. 
\newline
\newline
\textbf{Initial Search Instructions:} Imagine yourself in a history class. Your teacher has given you some time at the end of class to start your research for the assignment on the French Revolution. 
\newline
\newline
Since the class will be over in 15 minutes, you have decided that you would like to spend your time learning about the variety of information that is available about this topic. Your teacher has told you that you should not use web pages as your source of information, so you decide to search for books in the public library and save any you think will help you with your assignment. You have given yourself the goal of learning as much as you can about the topic, and to not make any decisions on the specific focus of your assignment. 
\newline
\newline
\textbf{Resumed Search Instructions:} About a week ago, you spent fifteen minutes in your history class looking for some resources to get you started with your assignment on the French Revolution. The French Revolution was a period of political and social upheaval in France from 1789 to 1799. The Revolution was triggered by a combination of factors, including a financial crisis, food shortages, and widespread dissatisfaction with the absolute monarchy and the privileges of the aristocracy.
\newline
\newline
Imagine you are now at home and have plenty of time to continue your previously unfinished search task of finding relevant resources for your assignment. Since your prior search activities focused on learning about the topic, your goal now is to pick up where you left off, and then focus on a specific aspect of the French Revolution so that your paper can have a narrow focus. Continue searching within the public library, save what you like, and delete those that are no longer relevant. 
    
\end{boxedminipage}
\caption{A simulated work task modelled after Borlund's model of simulated work tasks~\citep{Borlund97}.}
\label{fig:tasks}
\end{figure}

This simulated work task scenario aligns with the exploratory search model~\citep{White2009}, with the initial session representing an exploratory browsing phase, focused on learning about the topic. The second session shifts towards focused searching, emphasizing re-acquaintance with the previous work, further searching with a narrower focus, and the assessment of the relevance of collected documents.
        

\subsection{Participants}
A convenience sample of 32 participants were recruited from the undergraduate population at our university, with the first half assigned the Search Timelines interface and the second half assigned the Baseline. One participant assigned to the Baseline condition dropped out after the first session; the data for that participant was discarded and another participant was recruited. 

All participants provided responses to a demographic questionnaire at the outset of the initial session. The gender distribution among participants in this study consisted of 50\% females, 47\% males, and 3\% who identified as non-binary. Within this participant pool, demographic data on the level of education and English proficiency was collected. Participants were also asked how frequently they use public digital library search interfaces. There were no substantive differences between those who were assigned the Search Timelines interface versus the Baseline. The participants represented a broad sample of different education levels, with strong English language proficiency, and moderate use of the public digital library.

A side-effect of using a convenience sample was that we needed to situate the participants within a context of searching within the public library, rather than the university library. This was necessary because our data source was connected to a public library not an academic (university) library. We did this by asking the participants to imagine being a high school student, instead of a university student. This could have been avoided by recruiting among high school students; however doing so would have introduced complications associated with obtaining school board and research ethics approval. Using convenience samples and simulated search tasks were deemed an acceptable compromise. Of note, none of the participants expressed concerns with having to imagine being a high school student.



\subsection{Analysis}
With one independent variable, the data was analyzed using statistical comparisons between the 16 participants who used Search Timelines versus the 16 who used the Baseline. The survey data was aggregated per concept: aspects of and overall user engagement (RQ1a and RQ1b), aspects of usability (RQ2a and RQ2b), and perceived knowledge gain (RQ3a and RQ3b). The time to task completion data was used as-is (RQ4a/b). Shapiro-Wilk tests were used to assess whether the data was normally distributed. 

For each of the statistical comparisons of the survey data, at least one set of data was found to not be statistically significant. As a result, the non-parametric Mann-Whitney U test was used, along with Cohen's d for effect size measurements. A p-value of $0.05$ was used to determine statistical significance. Of note, when there are large differences in the data being analyzed, it is possible for the test statistic W to be the same value across many of the measures (in our case, 256). The effect size is interpreted as small when $d=0.2$, medium when $d=0.5$, and large when $d\ge0.8$.

For the statistical comparison of the time to task completion data, the distribution for each condition was normal. As a result, ANOVA tests were used along with eta-squared for effect size measurements. The effect size is interpreted as small when $\eta^2=0.01$, medium when $\eta^2=0.06$, and large when $\eta^2\ge0.14$.


\section{Results}\label{sec5}

The results from the user study are organized around the four pairs of research questions associated with the initial search and the resumed session.

\subsection{RQ1a: User Engagement During the Initial Search}
Participants using Search Timelines indicated higher levels of user engagement compared to those who used the Baseline, which were found to be statistically significant with medium to large effect sizes. The aggregated data for each condition, along with the statistical analysis are provided in Table~\ref{table:UES_Ini} for each aspect of user engagement, as well as overall.

\begin{table*}[!h]
\small
\begin{tabular}{l|llll}
  \begin{tabular}[c]{@{}l@{}}UES-SF Dimensions\\(Initial Search)\end{tabular}      & Baseline     & \begin{tabular}[c]{@{}l@{}}Search\\ Timelines \end{tabular}  & \begin{tabular}[c]{@{}l@{}} Mann-Whitney U\end{tabular}  & \begin{tabular}[c]{@{}l@{}}Effect Size\\ (Cohen's d) \end{tabular}   \\ \hline
  
  \begin{tabular}[c]{@{}l@{}}Focused Attention (FA)\end{tabular} &  2.79 (0.581) &  4.90 (0.541) & \textbf{W = 256, p \textless $\mathbf{10^{-6}}$} & d = 6.191  \\ \hline
  
   \begin{tabular}[c]{@{}l@{}} Perceived  Usability (PU) \end{tabular}     & 2.50 (1.13)  & 4.96 (0.462) & \textbf{W = 256, p \textless $\mathbf{10^{-6}}$} & d = 7.668   \\ \hline
        
   \begin{tabular}[c]{@{}l@{}}Aesthetic Appeal (AE)\end{tabular}        & 2.79 (0.713) &  4.69 (0.648) &  \textbf{W = 256, p \textless $\mathbf{10^{-5}}$} &  d = 6.03  \\ \hline
        
   \begin{tabular}[c]{@{}l@{}} Reward Factor (RW)\end{tabular}             & 2.62 (0.497) & 4.94 (0.454) & \textbf{W = 256, p \textless $\mathbf{10^{-6}}$} & d = 7.726  \\ \hline
        
   \begin{tabular}[c]{@{}l@{}}Overall User Engagement \end{tabular}        & 2.68 (0.262) & 4.87 (0.139) & \textbf{W = 256, p \textless $\mathbf{10^{-5}}$} & d = 10.464  \\ \hline
\end{tabular}
\caption{Mean values (standard deviation) for the user engagement scores collected after using the assigned search interface for the \textit{initial search} (data measured on 5-point Likert scales). 
}
\label{table:UES_Ini}
\end{table*}

\begin{table*}[!h]
\small
\begin{tabular}{l|llll}
  \begin{tabular}[c]{@{}l@{}}UES-SF Dimensions\\(Resumed Session)\end{tabular}      & Baseline     & \begin{tabular}[c]{@{}l@{}}Search\\Timelines \end{tabular}  & \begin{tabular}[c]{@{}l@{}} Mann-Whitney U\end{tabular}  & \begin{tabular}[c]{@{}l@{}}Effect Size\\ (Cohen's d) \end{tabular}      \\ \hline
  
  \begin{tabular}[c]{@{}l@{}}Focused Attention (FA)\end{tabular} & 2.50 (0.422) &  4.94 (0.181) & \textbf{W = 256, p \textless $\mathbf{10^{-6}}$}&  d = 7.511 \\ \hline
  
   \begin{tabular}[c]{@{}l@{}} Perceived Usability (PU) \end{tabular} & 2.35 (0.310) & 4.98 (0.083) & \textbf{W = 256, p \textless $\mathbf{10^{-6}}$} &  d = 11.579   \\ \hline
        
   \begin{tabular}[c]{@{}l@{}}Aesthetic Appeal (AE)\end{tabular}   & 2.77 (0.338) & 4.92 (0.149) & \textbf{W = 256, p \textless $\mathbf{10^{-6}}$} &  d = 8.212 \\ \hline
        
   \begin{tabular}[c]{@{}l@{}} Reward Factor (RW)\end{tabular} &2.40 (0.327) & 4.96 (0.167) & \textbf{W = 256, p \textless  $\mathbf{10^{-6}}$}&  d = 9.873  \\ \hline
        
   \begin{tabular}[c]{@{}l@{}}Overall User Engagement \end{tabular} & 2.51 (0.198) & 4.95 (0.079) & \textbf{W = 256, p \textless $\mathbf{10^{-5}}$}   &  d = 16.162 \\ \hline
\end{tabular}
\caption{Mean values (standard deviation) for the user engagement scores collected after using the assigned search interface in the \textit{resumed session} (data measured on 5-point Likert scales).
}
\label{table:UES_Res}
\end{table*}

We attribute this outcome to two main differences between Search Timelines and the Baseline: the high quality of the design and implementation of the new features in Search Timelines, contributing to perceived usability and aesthetic appeal measures; and the support that Search Timelines provides for tracking search activities such as queries issued and search results saved during the exploratory browsing process, contributing to focused attention and reward factor measures.

\subsection{RQ1b: User Engagement During the Resumed Session}
When resuming the search task, participants who used Search Timelines reported higher user engagement scores compared to those using the Baseline, mirroring the results observed during the initial session. These findings were statistically significant with medium to large effect sizes (see Table~\ref{table:UES_Res}).

This outcome affirms the value of providing the Detailed Timeline within the workspace, supporting the participants in reacquainting themselves with their previous search activities before they resumed the search activity. The quality of the workspace design contributed to measures of aesthetic appeal. Providing the timeline both within the workspace and as an overview within the SERP served as a visual reminder of what they had done as the participants continued their searches, contributing to focused attention, perceived usability, and reward factors.

\subsection{RQ2a: Usability During the Initial Search}
Participants who used Search Timelines also reported higher usability scores than those who used the Baseline, measured through aspects of ease-of-use, usefulness, and satisfaction. As shown in Table~\ref{table:U_ini}, these findings were statistically significant with medium to large effect sizes.

As both interfaces included the core abilities for searchers to issue queries, evaluate the search results, and save resources in a workspace, we attributed these differences to the Overview Timeline added to the SERP. As this timeline did not require the searcher to do anything different than their normal search process (issue queries and save search resources), participants reported that it was easy to use. Providing the timeline in a clear and logical visual structure contributed to its usefulness for keeping track of the search activity and overall satisfaction in the search process.

\begin{table*}[!h]
\small
\begin{tabular}{l|llll}
\begin{tabular}[c]{@{}l@{}}Usability Measure\\(Initial Search) \end{tabular}  & Baseline     & \begin{tabular}[c]{@{}l@{}}Search\\Timelines \end{tabular}  & \begin{tabular}[c]{@{}l@{}}Mann-Whitney U\end{tabular}  &     \begin{tabular}[c]{@{}l@{}}Effect Size\\ (Cohen's d) \end{tabular}    \\ \hline

Ease-of-use &  2.67 (0.472) &  4.95 (0.101) &  \textbf{W = 256, p \textless  $\mathbf{10^{-6}}$} &  d = 6.686 \\ \hline

Usefulness & 2.47 (0.386)  & 4.94 (0.144) & \textbf{W = 256, p \textless $\mathbf{10^{-6}}$} &  d = 8.473 \\ \hline

Satisfaction & 2.45 (0.306) & 4.92 (0.198) & \textbf{W = 256, p \textless $\mathbf{10^{-6}}$} &  d = 9.580 \\ \hline
\end{tabular}
\caption{Mean values (standard deviation) for the usability measure collected after using the assigned search interface in the \textit{initial search} (data measured on 5-point Likert scales).
}
\label{table:U_ini}
\end{table*}

\begin{table*}[!h]
\small
\begin{tabular}{l|llll}
\begin{tabular}[c]{@{}l@{}}Usability Measure\\(Resumed Session) \end{tabular}  & Baseline     & \begin{tabular}[c]{@{}l@{}}Search\\Timelines \end{tabular}  & \begin{tabular}[c]{@{}l@{}}Mann-Whitney U\end{tabular}  &     \begin{tabular}[c]{@{}l@{}}Effect Size\\ (Cohen's d) \end{tabular}    \\ \hline

Ease-of-use &  2.59 (0.364) &  4.97 (0.085) &  \textbf{W = 256, p \textless $\mathbf{10^{-6}}$} &  d = 8.990 \\ \hline

Usefulness & 2.20 (0.332)  & 4.95 (0.188) & \textbf{W = 256, p \textless $\mathbf{10^{-6}}$} &  d =  10.202 \\ \hline

Satisfaction & 2.58 (0.285) & 4.92 (0.176) & \textbf{W = 256, p \textless $\mathbf{10^{-6}}$} &  d = 9.905 \\ \hline
\end{tabular}
\caption{Mean values (standard deviation) for the usability measures collected after using the assigned search interface in the \textit{resumed session} (data measured on 5-point Likert scales).
}
\label{table:U_res}
\end{table*}

\subsection{RQ2b: Usability During the Resumed Session}

In the resumed session, participants who used Search Timelines continued to reporting more positively about the usability compared to those who used the Baseline. These differences were found to be statistically significant with large effect sizes (see Table~\ref{table:U_res}).

The key feature provided to support resuming the search task was the Detailed Timeline within the workspace. Replicating the search result cards from the SERP into the timeline resulted in a familiar structure in the interface, which contributed to its ease of use. The extra contextual information that linked each saved resource to its source query and temporal information, contributed to perceptions of usefulness and satisfaction. These aspects of usability were further supported by providing the timeline in an overview format when the searchers returned to issuing queries and viewing search results.

\subsection{RQ3a: Perceived Knowledge Gain During the Initial Search}
The differences in the perceived knowledge measured pre-search and post-search during the initial search session were calculated separately for those who used the Baseline and those who used Search Timelines. For both cases, we found an increase in the perceived knowledge that was statistically significant with small effect sizes (see Table~\ref{table:KG_ini}). This finding shows that the participants of both interfaces were actively engaged in the search process and found search results that felt enhanced their knowledge during the initial search.

We also conducted a statistical analysis comparing the degree of knowledge change between the two interface conditions. Statistical significance with a small effect size was found in favour of Search Timelines ($W = 245.5$, \textbf{p \textless $\mathbf{10^{-5}}$}, $d = 2.365$). This outcome highlights the value of providing a visual timeline of queries and saved resources within the SERP, and its positive effect on supporting the learning that is critical in the early stages of exploratory search.

\begin{table*}[!h]
\small
\begin{tabular}{l|lllll}
\multicolumn{1}{c|}{\begin{tabular}[c]{@{}l@{}}Interface\\ (Initial Search)\end{tabular}} &
  \begin{tabular}[c]{@{}l@{}}Perceived\\ Knowledge\\ (Pre-Search)\end{tabular} &
  \begin{tabular}[c]{@{}l@{}}Perceived\\ Knowledge\\ (Post-Search)\end{tabular} &
  $\nabla$ &
  \begin{tabular}[c]{@{}l@{}}Paired Mann-Whitney U\end{tabular} &
  \begin{tabular}[c]{@{}l@{}}Effect Size\\ (Cohen's d)\end{tabular} \\ \hline
Baseline &
  2.38 (0.631) &
  3.17 (0.544) &
  +0.79 &
  \textbf{V = 0, p \textless 0.001} &
  d = 1.343  \\
Search Timelines &
  2.54 (0.607) &
  4.65 (0.257) &
  +2.11 &
  \textbf{V = 0, p \textless 0.0005} &
  d = 4.513 \\ \hline
\end{tabular}
\caption{Mean values (standard deviation) for the perceived knowledge gain collected before and after using the search interfaces in the \textit{initial search session} (data measured on 5-point Likert scales).}
\label{table:KG_ini}
\end{table*}

\begin{table*}[!h]
\small
\begin{tabular}{l|lllll}
\multicolumn{1}{c|}{\begin{tabular}[c]{@{}l@{}}Interface\\ (Resumed Session)\end{tabular}} &
  \begin{tabular}[c]{@{}l@{}}Perceived\\ Knowledge\\ (Pre-Search)\end{tabular} &
  \begin{tabular}[c]{@{}l@{}}Perceived\\ Knowledge\\ (Post-Search)\end{tabular} &
  $\nabla$ &
  \begin{tabular}[c]{@{}l@{}}Paired Mann-Whitney U\end{tabular} &
  \begin{tabular}[c]{@{}l@{}}Effect Size\\ (Cohen's d)\end{tabular} \\ \hline
Baseline &
  2.54 (0.515) &
  2.69 (0.537) &
  +0.15 &
  V = 12, p = 0.2328 &
      \\
Search Timelines &
  3.29 (0.167) &
  4.88 (0.342) &
  +1.59 &
  \textbf{V = 0, p \textless 0.0005} &
  d = 5.892   \\ \hline
\end{tabular}
\caption{Mean values (standard deviation) for the perceived knowledge gain collected before and after using the search interfaces in the \textit{resumed search session} (data measured on 5-point Likert scales).}
\label{table:KG_res}
\end{table*}

\subsection{RQ3b: Perceived Knowledge Gain During the Resumed Session}
For the resumed search session, the degree of perceived knowledge gain was not as substantial as what was measured during the initial search session; the gain was only statistically significant for the participants who used Search Timelines, yet still with a medium effect size (see Table~\ref{table:KG_res}). When comparing the perceived knowledge gain between the interfaces, the increase for those who used Search Timelines was greater than for those who used the Baseline, at a statistically significant level and with a small effect size ($W = 253$, \textbf{p \textless $\mathbf{10^{-5}}$}, $d = 3.692$).

These findings showcase the challenges that the participants who used the Baseline experienced when resuming their search tasks. There was little perceived retention of the previous search activities, and with the typical digital library search interface and simple workspace, they had difficulty focusing their resumed search to the point where their perception of knowledge would increase.

In contrast, the participants who used Search Timelines had a notably higher perceived knowledge level when starting the resumed session compared to when they started the initial search. This is evidence of better knowledge retention during the time interval between searches, which we attribute to the visualization of the timeline. That is, the memory having been able to see the queries and associated saved resources in a structured graphical format, the perception of knowledge remained high. 

During the resumed session, participants who used Search Timelines were able to build upon their retained knowledge, reacquaint themselves with their prior work, and continue to learn about the topic. Clearly, the visual structure provided by the timelines (both in the workspace and in the SERP) helped the participants to develop confidence in their level of knowledge on the assigned topic.


\subsection{RQ4a: Time to Task Completion During the Initial Search}
Although the time available for the initial search was limited to 15 minutes (900 seconds) in order to create a need for cross-session searching, all participants completed their initial search activities before this time limit expired (see Table~\ref{tab:time-initial}). 


Those who used Search Timelines spent a statistically significant longer time (more than two minutes) conducting the initial search, with a large effect size. Their willingness to search longer may have been influenced by the presence of the timeline within the SERP, providing confidence that their search activities were being saved and that having to stop in the middle of a search because the time had expired would not be a problem. Conversely, those who used the Baseline stopped their searching with more than half the time remaining, showing that they had more difficulty self-monitoring their search activities and erring on the side of caution rather than running the risk of having to stop mid-search.

\begin{table*}[!h]
\begin{tabular}{l|llll}
\begin{tabular}[c]{@{}l@{}}Performance Measure \\ (Initial Search)\end{tabular} &
Baseline & \begin{tabular}[c]{@{}l@{}}Search\\Timelines \end{tabular}      & ANOVA   & \begin{tabular}[c]{@{}l@{}}Effect Size\\ (eta-squared) \end{tabular}     \\ \hline
\begin{tabular}[c]{@{}l@{}}Time to Task Completion\end{tabular}  & 400 (87.8)  & 533 (94.5)  & \begin{tabular}[c]{@{}l@{}}\textbf{F(1, 32) = 17.03,} \textbf{p \textless 0.0005}  \end{tabular} & $\eta^{2} = 0.362$\\ \hline

\end{tabular}
\caption{Mean values (standard deviation) of the time (seconds) spent in both interfaces during the initial session.
}
\label{tab:time-initial}
\end{table*}

\begin{table*}[!h]
\begin{tabular}{l|llll}
\begin{tabular}[c]{@{}l@{}}Performance Measure \\ (Resumed Session)\end{tabular} &
Baseline & \begin{tabular}[c]{@{}l@{}}Search\\Timelines \end{tabular}      & ANOVA   & \begin{tabular}[c]{@{}l@{}}Effect Size\\(eta-squared) \end{tabular}     \\ \hline
\begin{tabular}[c]{@{}l@{}}Time to Task Completion\end{tabular} & 467 (54.0)  & 869 (90.2)  & \begin{tabular}[c]{@{}l@{}}\textbf{F(1, 32) = 233.3,} \textbf{p \textless $\mathbf{10^{-14}}$}  \end{tabular} & $\eta^{2} = 0.886$\\ \hline

\end{tabular}
\caption{Mean values (standard deviation) of the time (seconds) spent in both interfaces during the resumed session.
}
\label{tab:time-resumed}
\end{table*}

\subsection{RQ4b: Time to Task Completion During the Resumed Session}

Although the resumed search task did not have a time limit, participants who used the Baseline interface did not take much more time in this session than they had in the initial search (about one more minute); by contrast, participants who used Search Timelines took over five more minutes in the resumed session (see Table~\ref{tab:time-resumed}). There was a statistically significant difference with a large effect size in the time taken between these two groups of participants.

While typical information retrieval studies identify a longer time as a negative result, we consider it a sign of engagement in pursuing the complex search task. As the scenario presented to the participants was one of conducting an exploratory search, being faster is not an expected goal; what is expected is that the searchers engage in the search process. The tendency of the participants who used the Baseline to end their resumed search sessions quickly is consistent with their low user engagement scores, usability scores, perceived knowledge gain.

Further, any time something new is added to a search interface, we can expect that there will be a penalty to be paid in terms of the time it takes a searcher to view the information provided, make sense of it in relation to the search activity, and make use of the interactive elements. Although we were not able to isolate the time it took searchers to use the new features from the time they spent engaging in the search process, the positive subjective results (user engagement and usability) suggest that this extra time taken was not viewed negatively.

\section{Discussion}\label{sec12}

The significant differences in the measures between the two conditions in this study warrant further discussion. Participants who used the Baseline reported rather negative opinions regarding user engagement and usability, had small increases in perceived knowledge after the search sessions, and finished their search activities quickly. Since the interface they used was closely modelled after a typical public digital library search interface, this shows that such search interfaces are inadequate for cross-session exploratory search. Conversely, participants who used Search Timelines reported very positive opinions regarding user engagement and usability, had substantial increased in perceived knowledge after the search sessions, and took their time with the search activities. This highlights the benefits of providing visual representations of the search activities when undertaking cross-session exploratory search. Of note, while we cannot determine whether the high user engagement lead to longer times to task completion or the other way around, what we can surmise is that together, these gave the participants the opportunity to engage with the search activity, resulting in their preceived knowledge gain.

Since we used a between-subjects design, neither participant group saw the other interface, although those who used Search Timelines were familiar with the public digital library search interface upon which the Baseline was modelled. As such, it is possible that the high user engagement and usability scores reported by the participants who used Search Timelines were influenced by their ability to make comparisons to the existing public library search interface. While this is an unavoidable side-effect of conducting a study that compares a new approach to a well-established baseline, it increases our confidence in the ecological validity of the findings given the consistency  with the perceived knowledge gain and the time taken to pursue the search task.

While there are a variety of reasons that a search might span multiple sessions~\citep{Capra20}, few search interfaces provide explicit support for assisting searchers in reacquainting themselves with past search activities as they resume their search. The results from this study are aligned with similar findings on the value of providing explicit support for cross-session searching~\citep{Gomes-Hoeber-22}.

These results are also aligned with prior work on the value of adding information visualization elements to search interfaces~\citep{Hoeber18, hoeber-2025}. Given that search interfaces are already very text-heavy, anything that can be done to help searchers see relationships within the data will be better than providing more information that needs to be read. In this work, the visualizations are subtle, but the designs follow established information visualization principles (e.g., the Gestalt Principles) that have a basis in innate human information processing abilities. Further, the choice to provide a lightweight visualization for the SERP and a more detailed view of the saved resources in the workspace provided what we considered the right amount of information detail for these two views (overview for the SERP, detail for the workspace) while maintaining the contextual information necessary to show the history of the search activity.

This study provides further evidence that the ``query box and ten blue links'' search interface design pattern is insufficient when search tasks become complex and there is a need to engage in cross-session exploratory search. We have shown that providing the context under which resources are saved to a workspace increases subjective opinion regarding user engagement and usability. We have also shown that providing access to this information also encourages learning, as measured through perceived knowledge gain. While these results are at the expense of being efficient, efficiency in an exploratory search is not a goal that should be encouraged. Exploratory search inherently requires more time for investigation and learning activities. The extended time spent indicates that participants were actively engaged and willing to delve deep into their search activities. Instead of holding objective performance as the ultimate criterion for success, we ascribe to Belkin's view on interactive information retrieval, where usefulness should be the main evaluation criterion~\citep{Belkin2016}.

A particularly interesting finding from this study was the retention of perceived knowledge between the search sessions for those using Search Timelines, which was not present for those who used the Baseline. This is evidence that providing a visual representation of search activities within the SERP can enhance the perception of knowledge retention. Further study of different types of search activity summaries that vary the degree of visual/textual information are warranted, along with studies of whether actual knowledge retention follows what we have found with perceived knowledge retention.

Although we found rather significant differences between the participants who used the two interfaces, with many medium to large effect sizes, the actual differences between the interfaces was only the inclusion of the timelines. All other information and interaction mechanisms were the same (issuing queries, viewing search results, saving resources to the workspace). Given this, along with the between-subjects design that resulted in each participant using only one of the two interfaces, we have high confidence in the validity of these findings. While the participants assigned to the Baseline condition had a slightly lower initial perceived knowledge level (see Table~\ref{table:KG_ini}), this difference of 0.16 was due to random chance and not statistically significant. 

One limitation of this study was the use of a convenience sample of university students, who were then asked them to imagine being high school students and users of the public library. We can expect that university students have developed search skills through their studies, and therefore would perform better in the search task than the average public library patron. Further study with a broad sample of public library patrons with different levels of cognitive ability, search expertise, digital literacy, and complex search task needs would provide evidence of whether this approach is more or less valuable for different subsets of the public library patron population. 

Another limitation was that measurements of knowledge gain were based on the perceptions of the participants rather than their actual knowledge about the topic. Future studies that measure both perceived and actual knowledge will provide a more complete picture of how this approach can support knowledge gain.

\section{Conclusion}\label{sec13}

The primary contributions of this research are the design of the Search Timelines interface which uses visualization techniques to support cross-session exploratory search, a prototype implementation of the design, and the results of a user study that provide evidence of the benefits of this approach in the context of public library search activities. This work represents an example of how lightweight information visualization techniques that are scrutable, interactive, and persistent can be used to enhance interactive information retrieval processes~\citep{hoeber-2025}.

The approach was developed to provide timelines at two levels of detail (overview in the SERP and full detail in the workspace). It may be possible to design even more compact timeline visualizations that abstract away even more details and use interactive methods such as the hover operation to reveal such details on demand. Similarly, it may be possible to design even more detailed timelines that add additional information on each saved resource in the workspace. Further study may reveal what is the optimal amount of information to provide searchers to support task resumption (in the workspace) and monitoring of search activities (in the SERP). 

The study was designed within the specific context of searching within public digital libraries. Porting the approach to other search contexts where exploratory search is prevalent (e.g., academic digital libraries or searching within digital humanities archives) and conducting further user studies will add evidence regarding the generalizability of the approach. Extending these studies into naturalistic settings will address ecological validity limitations. Additional studies that consider and collect evidence regarding the knowledge retention we observed in this study are warranted, as are studies that explore the benefit of the approach from the context of metacognitive planning and monitoring~\citep{schraw_metacognitive_1998}.

\backmatter


\bibliography{references}

\end{document}